\documentclass{icrc29}
\usepackage{graphicx,amssymb,amsmath,times}
\usepackage{overpic}
\usepackage{psfrag}
\usepackage[rflt]{floatflt} 

\DeclareGraphicsExtensions{.eps,.ps,.eps.gz,.ps.gz}
\begin{document}

\title[The Ring Imaging Cherenkov detector (RICH) of the AMS experiment]{The Ring Imaging Cherenkov detector (RICH) of the AMS experiment}
\author[F. Barao, for the AMS RICH collaboration] {%
       M. Aguilar Benitez$^1$, J. Alcaraz$^1$, L. Arruda$^2$, F. Barao$^2$, 
       A. Barrau$^3$,  G. Barreira$^2$, E. Belmont$^4$, \newauthor
       J. Berdugo$^1$, M. Brinet$^3$, M. Buenerd$^3$, D. Casadei$^5$, J. Casaus$^1$, E. Cortina$^1$, 
       C. Delgado$^6$, \newauthor
       C. Diaz$^1$, 
       L. Derome$^3$, L. Eraud$^3$, 
       R.J. Garcia-Lopez$^6$, L. Gallin-Martel$^3$, F. Giovacchini$^5$, \newauthor
       P. Goncalves$^2$, 
       E. Lanciotti$^1$, G. Laurenti$^5$, A. Malinine$^7$, C. Mana$^1$, 
       J. Marin$^1$, G. Martinez$^1$, \newauthor 
       A. Menchaca-Rocha$^4$, 
       M. Molla$^1$, C. Palomares$^1$,  M. Panniello$^6$, R. Pereira$^2$, M. Pimenta$^2$, \newauthor 
       K. Protasov$^3$, E. Sanchez$^1$, 
       E-S. Seo$^7$, N. Sevilla$^1$, A. Torrento$^1$, M. Vargas-Trevino$^3$, O. Veziant$^3$ \\
      (1) CIEMAT, Avenida Complutense 22, E-28040, Madrid, Spain \\
      (2) LIP, Avenida Elias Garcia 14-1, P - 1000 Lisboa, Portugal \\
      (3) LPSC, Avenue des Martyrs 53, F-38026 Grenoble-cedex, France \\
      (4) Instituto de Fisica, UNAM, AP 20-364, Mexico DF, Mexico \\
      (5) University of Bologna and INFN, Via Irnerio 46, I-40126 Bologna, Italy \\
      (6) IAC, C/Via Lactea s/n, E-38200, La Laguna, Tenerife, Spain \\
      (7) U. Maryland, College Park MD 20742, USA 
}
\presenter{Presenter: F. Barao (barao@lip.pt), \
por-barao-F-abs1-og15-oral }

\maketitle

\begin{abstract}

The Alpha Magnetic Spectrometer (AMS) experiment to be installed
on the International Space Station (ISS) will be equipped with a
proximity focusing Ring Imaging Cherenkov (RICH) detector
for measuring the electric charge and velocity of the charged 
cosmic particles.
A RICH prototype consisting of 96 photomultiplier units, including a piece 
of the conical reflector, was built
and its performance evaluated with ion beam data. 
Preliminary results of the in-beam tests performed with ion fragments 
resulting from collisions of a 158 GeV/c/nuc primary beam of Indium ions (CERN SPS) 
on a Pb target are reported. 
The collected data included tests to the 
final front-end electronics and to different aerogel radiators. 
Cherenkov rings for a large range of charged 
nuclei and with reflected photons were observed.
The data analysis confirms the design goals. 
Charge separation up to Fe 
and velocity resolution of the order of 0.1\% for singly charged particles are
obtained. 

\end{abstract}

\vspace{-0.5cm}
\section{Introduction}
\vspace{-0.2cm}
The Alpha Magnetic Spectrometer~\cite{bib:ams} (AMS) is a high energy physics experiment that 
will be installed on the International Space Station (ISS) by the year 2008, where 
it will operate for a period of at least three years. 
It is a large acceptance (\textrm{$\sim 0.5$ m$^2$sr}) superconducting magnetic spectrometer
able to detect in a wide kinematic range (from a few hundred MeV up to TeV region) singly charged 
particles, charged nuclei 
and $\gamma$ rays.
The long time exposure in space will allow AMS to collect an unprecedented large
data sample and to extend by orders of magnitude the sensitivity reached by previous 
experiments on dark matter and antimatter searches.
In addition, the measurement of the cosmic-ray abundances up to the TV region and in a wide 
charge range (up to Z$\sim 26$) will contribute to a better description of cosmic ray production,
acceleration and propagation mechanisms, essential for a full understanding of the background spectra 
on dark matter searches.
Information about the density of the interstellar medium traversed by the 
cosmic rays and their confinement time can be derived from the isotopic composition of 
secondary cosmic rays, produced by fragmentation during the cosmic ray transport in the galaxy.
For instance, the relative abundances of deuterium and helium-3 isotopes reflect the transport 
history along the galaxy of protons and heliums, while the beryllium-10 radionuclide  accounts for 
the time confinement.      
Current measurements are performed at relatively low energies (T $\lesssim 1$ GeV/n ) and based on 
small statistics. 

\begin{figure}
\parbox{0.45\textwidth}{%
 \begin{overpic}[width=.4\textwidth]{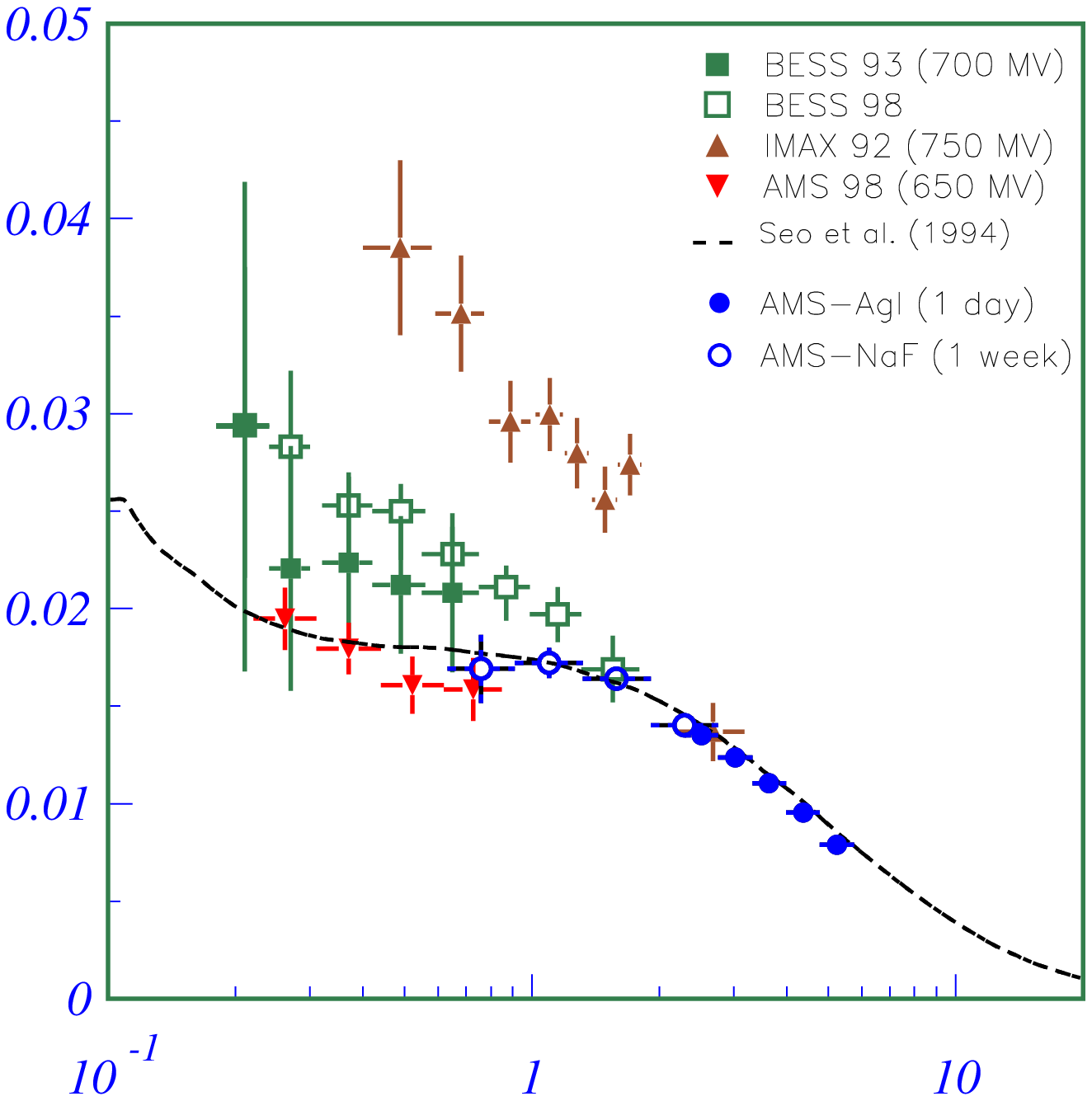}
 \put(17,80){{$^2H/^1H$}}
 \put(35,0){{Kinetic energy (GeV/n)}}
 \end{overpic}
}
\parbox{0.45\textwidth}{\caption{%
AMS expected isotopic ratio for $^2H/^1H$ (open and full dots) as compared with AMS01 measurements
obtained during a shuttle flight in 1998 and two balloon experiments.
The expected results are based on simulated data equivalent to 1 week statistics for events 
passing through the RICH central radiator (NaF) and 1 day for events passing elsewhere.
The events were simulated according to the reacceleration model of reference~\cite{bib:crays-seo1994} 
represented by the dashed line.
}
}
\label{fig:ams02-sim-dpratio} 
\end{figure}

Particle identification with AMS-02 relies on a very precise determination 
of the magnetic rigidity, energy, velocity and electric charge.
In the AMS spectrometer, the momentum is obtained from the information  
provided by the silicon tracker
with a relative  accuracy of $\sim$1\% up to $10$ GeV/c/n. 
Isotopic mass separation over a wide range of energies requires,
in addition to an accurate momentum measurement, 
a velocity determination with low relative uncertainty in 
as 
\mbox{$\Delta m/m = ( \Delta p / p ) \oplus \gamma^2 ( \Delta \beta / \beta )$}.
For this purpose, the AMS spectrometer includes a Ring Imaging Cherenkov detector (RICH) 
operating between the time-of-flight and electromagnetic calorimeter (ECAL) detectors.
It was designed to provide measurements of the velocity for singly charged particles with a 
relative uncertainty of 0.1\% and of the nuclei electric charge up to Fe.
Moreover, it will provide AMS with an additional contribution to the electron/proton separation.  
For the isotopic separation, the RICH detector will cover a kinetic energy region ranging from 
0.5 GeV/n up to around 10 GeV/n for A $\lesssim 10$.
Figure~\ref{fig:ams02-sim-dpratio} shows the expected isotopic deuterium-proton ratio to be measured 
by AMS, based on a simulated data sample of $\sim 10^7$ events~\cite{bib:lip-isotopes}.
Although there is an upper boundary around 6 GeV/n, imposed by the low fraction of deuterium signal in comparison 
with the dominant proton mass tail ($d/p \sim 10^{-2}$), the current kinematic region is clearly extended.

\vspace{-0.5cm}
\section{The AMS RICH detector}
\vspace{-0.2cm}
The RICH design was driven by a set of constraints imposed by the 
launch and the long duration flight environment on one hand, and by the integration in AMS and the 
envisaged physics aims, on the other hand.
Therefore, the RICH options had to deal with restrictions on size, weight, power consumption and 
materials. In addition, it had to take into account the AMS stray magnetic field, reaching $\sim 300$ G 
in some locations, and the minimization of matter in front of the electromagnetic calorimeter.

\begin{figure}
 \begin{minipage}[b]{.64\linewidth}
   \begin{tabular}{cc}
     \includegraphics*[width=0.5\textwidth,angle=0,clip]{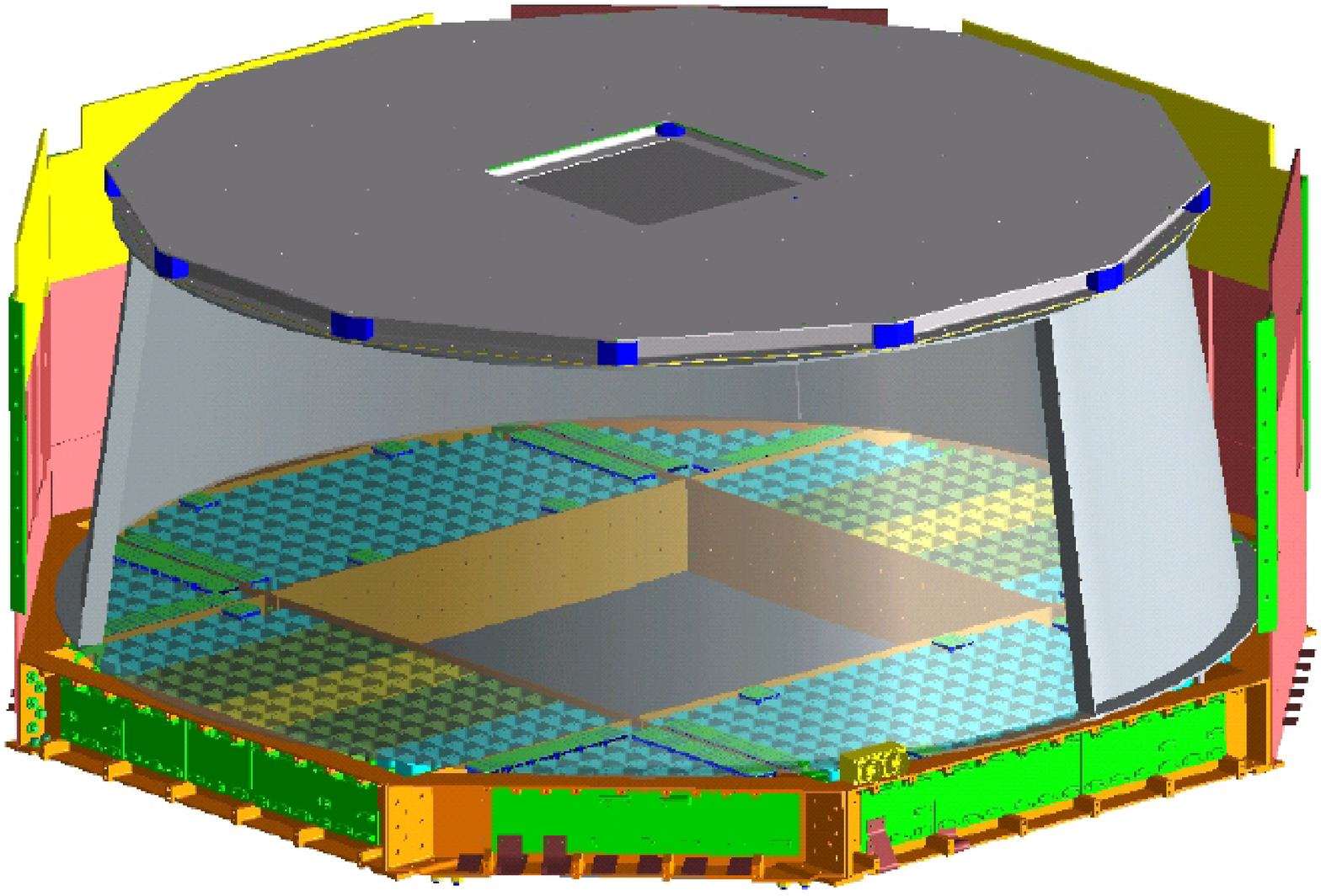}  & 
     \includegraphics*[width=0.45\textwidth,angle=0,clip]{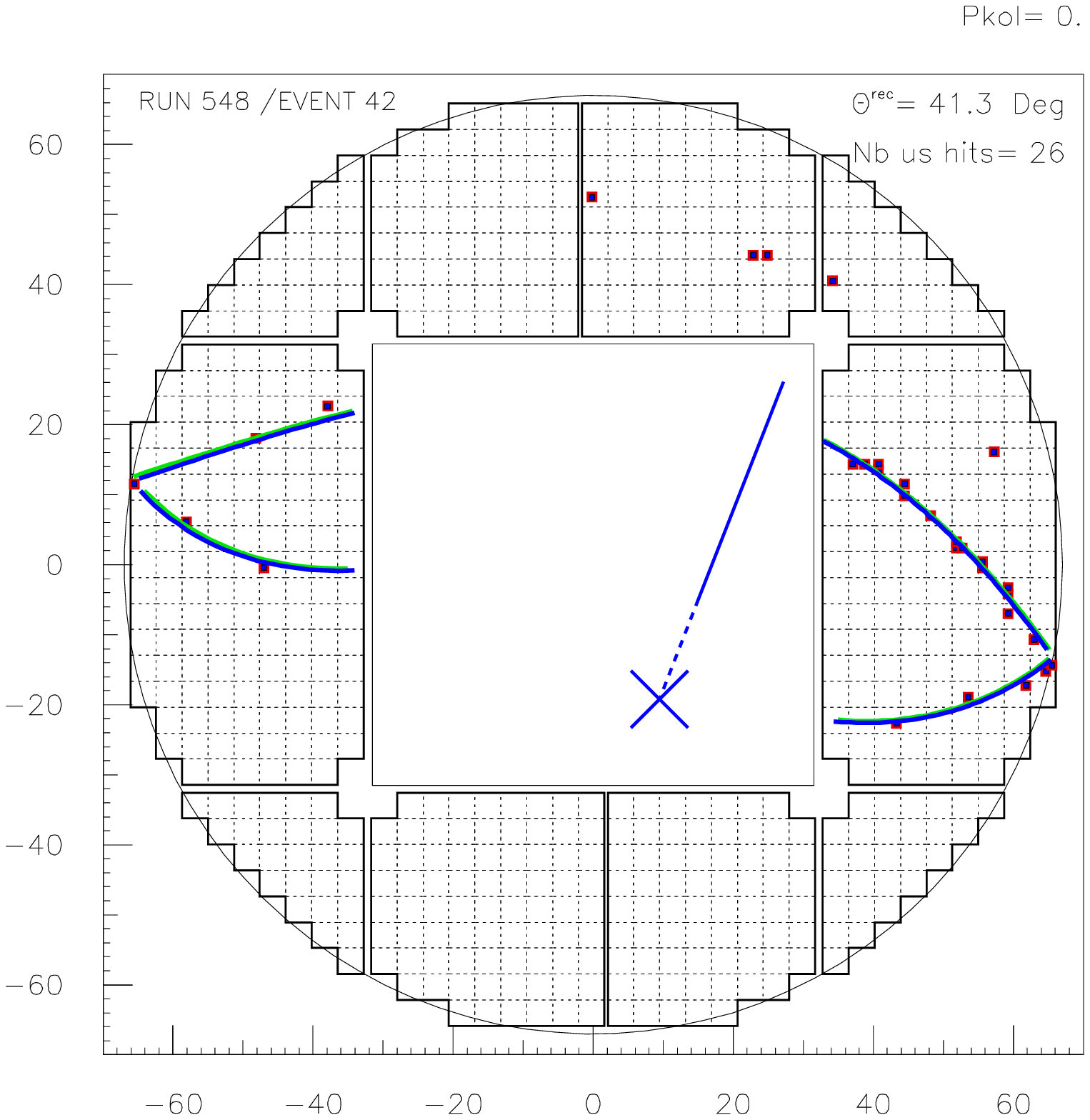} 
   \end{tabular}
  \caption{Schematic view of the RICH detector.
           Display of a reconstructed photon ring for a simulated beryllium event crossing the NaF radiator.
    \label{fig:rich-3D}}
 \end{minipage} \hfill
 \begin{minipage}[b]{.34\linewidth}
   \centering\includegraphics*[width=0.8\textwidth,angle=0,clip]{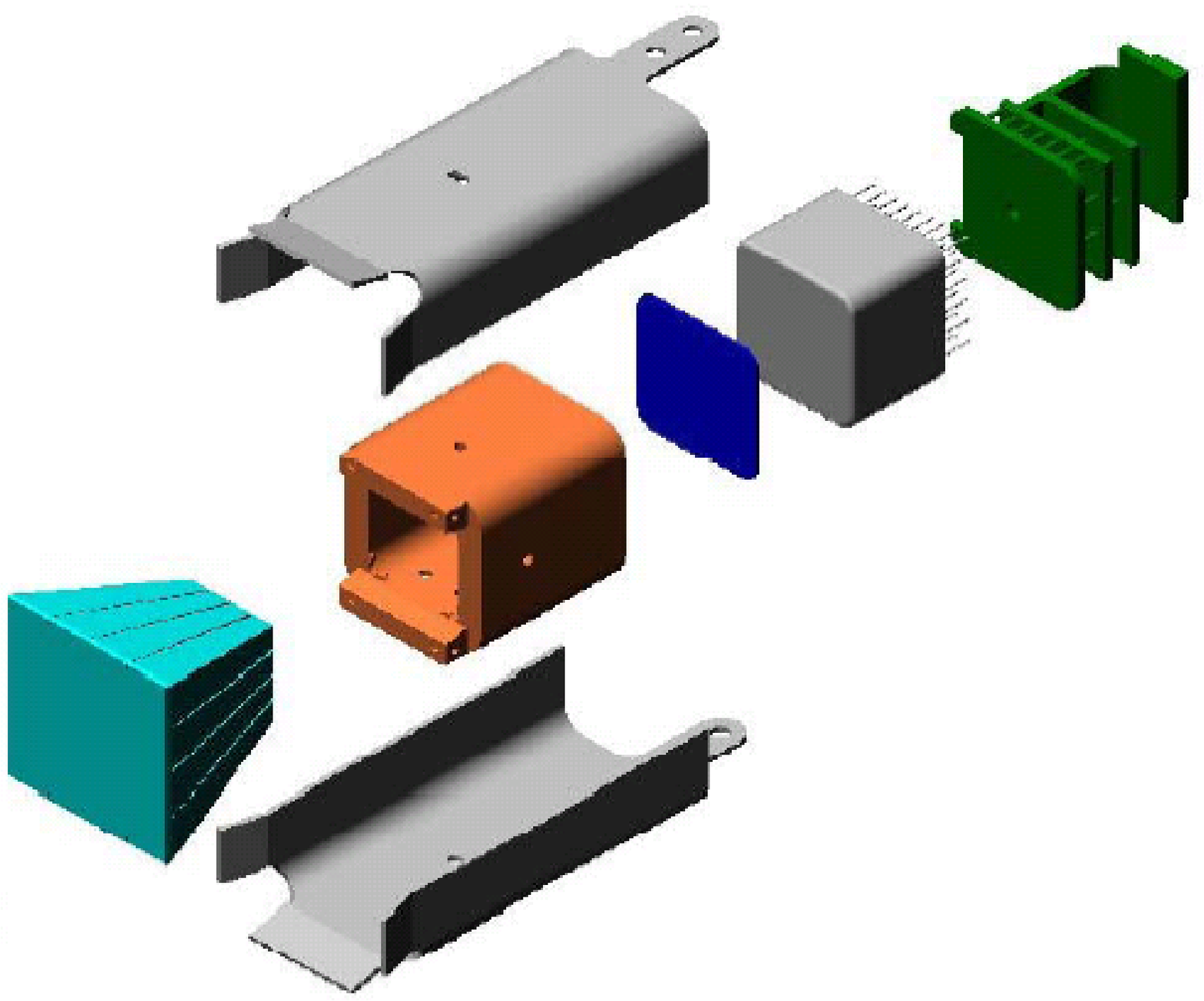} 
   \caption{Detailed view of a readout cell. 
     \label{fig:rich-unitcell}}
 \end{minipage}
\end{figure}

The RICH has a truncated conical shape with a top radius of 60 cm, a bottom radius of 67 cm, and a total
height of 60.5 cm. It covers 80\% of the AMS magnet acceptance~\cite{bib:ams02-rich}.
A general view of the RICH detector is shown in figure~\ref{fig:rich-3D}.
It is a proximity focusing device with a dual solid radiator configuration on the top, 
an expansion height of 46.9 cm and, at the bottom, a matrix of 680 multipixelized photon readout cells.
A high reflectivity mirror with a conical shape surrounds the whole set in order to increase the
device acceptance.  
The radiator is made of $92$ aerogel $27$ mm thick tiles with a refractive index 1.05,  and 
sodium fluoride (NaF) tiles with a  thickness of $5$ mm in the center
covering an area of $34\times34$ cm$^2$.  
The NaF placement prevents the loss of photons in the hole existing in the center of the readout plane  
($64\times64$ cm$^2$), in front of the ECAL calorimeter located below.
Figure~\ref{fig:rich-3D} shows a NaF  event display.
The radiator tiles are supported by a $1$ mm thick layer of methacrylate (n=$1.5$) free of UV absorbing
additives.


To prevent a large fraction of RICH radiated photons ($\sim 33$\%) to escape through the lateral surface 
of the expansion volume, a conical reflector was designed. 
It consists of a carbon fiber reinforced composite substrate with a multilayer coating made of
aluminium and SiO$_2$ deposited on the inner surface. 
This ensures a reflectivity higher than 85\% for $420$ nm wavelength photons. 

The photon detection is made with an array of multianode Hamamatsu tubes (R7600-00-M16) with a spectral 
response ranging from $300$ to $650$ nm and a maximum at $\lambda \sim$ 420 nm.
The choice of the phototube was driven, among other factors, by its response to the photoelectron signal and
its low sensitivity to the magnetic field.
Nevertheless, the strength of the residual field from the superconducting magnet imposes the need to shield the 
photomultipliers with a permalloy thickness varying from 0.8 to 1.3 mm. 
To increase the photon collection efficiency, a light guide consisting of 16 solid acrylic pipes 
glued to a thin top layer ($1$ mm) was produced. 
It is optically coupled to the active area of phototube cathode through a $1$ mm flexible optical pad.
With a total height of $31$ mm and a collecting surface of $34\times34$ mm$^2$, it presents a readout pixel size 
of $8.5$ mm.  
The light guide is mechanically attached through nylon wires to the photomultiplier polycarbonate housing.

The detected photons are converted into a charge signal in the photomultiplier with a typical gain of $\sim 10^6$.
A low consumption 80 M$\Omega$ high voltage divider was chosen.
The charged signal is then shaped and amplified ($\times 1$ or $\times 5$) in a front-end chip in order to both 
cope  with a dynamic range of $10^2$ and to keep a high sensitivity to the photoelectron signal.    
Finally the signal is digitized on a 12-bit ADC.
The RICH data acquisition system  deals with a total number of $10,880$ readout channels.
Figure~\ref{fig:rich-unitcell} shows an exploded view of a complete readout cell with all the chain from the light 
guide to the front-end electronics.

RICH assembling activities started in September 2003. The final detector is scheduled to be operational at the 
beginning of 2006 for functionality tests and further integration into AMS. 

\vspace{-0.5cm}
\section{The RICH prototype}
\vspace{-0.2cm}
A prototype of the RICH detector consisting of an array of $9 \times 11$ cells filled with 96 photomultiplier readout 
units was constructed. Its performance was evaluated with cosmic muons and fragmented ions 
from CERN SPS beams in 2002~\cite{bib:ams02-rich} and 2003. 
The light guides used were prototypes with a slightly smaller size ($31$ mm).
An adjustable supporting structure was used to test different sets of aerogels at variable expansion heights.
The setup was completed with AMS silicon tracker layers placed upstream in the beam, two multi-wire proportional 
chambers and scintillator counters.  
Secondary fragments with charges $Z<49$ from the fragmentation of a $158$ GeV/c indium beam were used 
in the 2003 run. 
Given the small angular acceptance of the beam line, a rigidity accuracy of 1.5\% was provided. 
A total number of 11 million events were recorded during seven days. 

\begin{figure}
 \begin{minipage}[b]{.32\linewidth}
   \includegraphics*[width=0.95\textwidth,angle=0,clip]{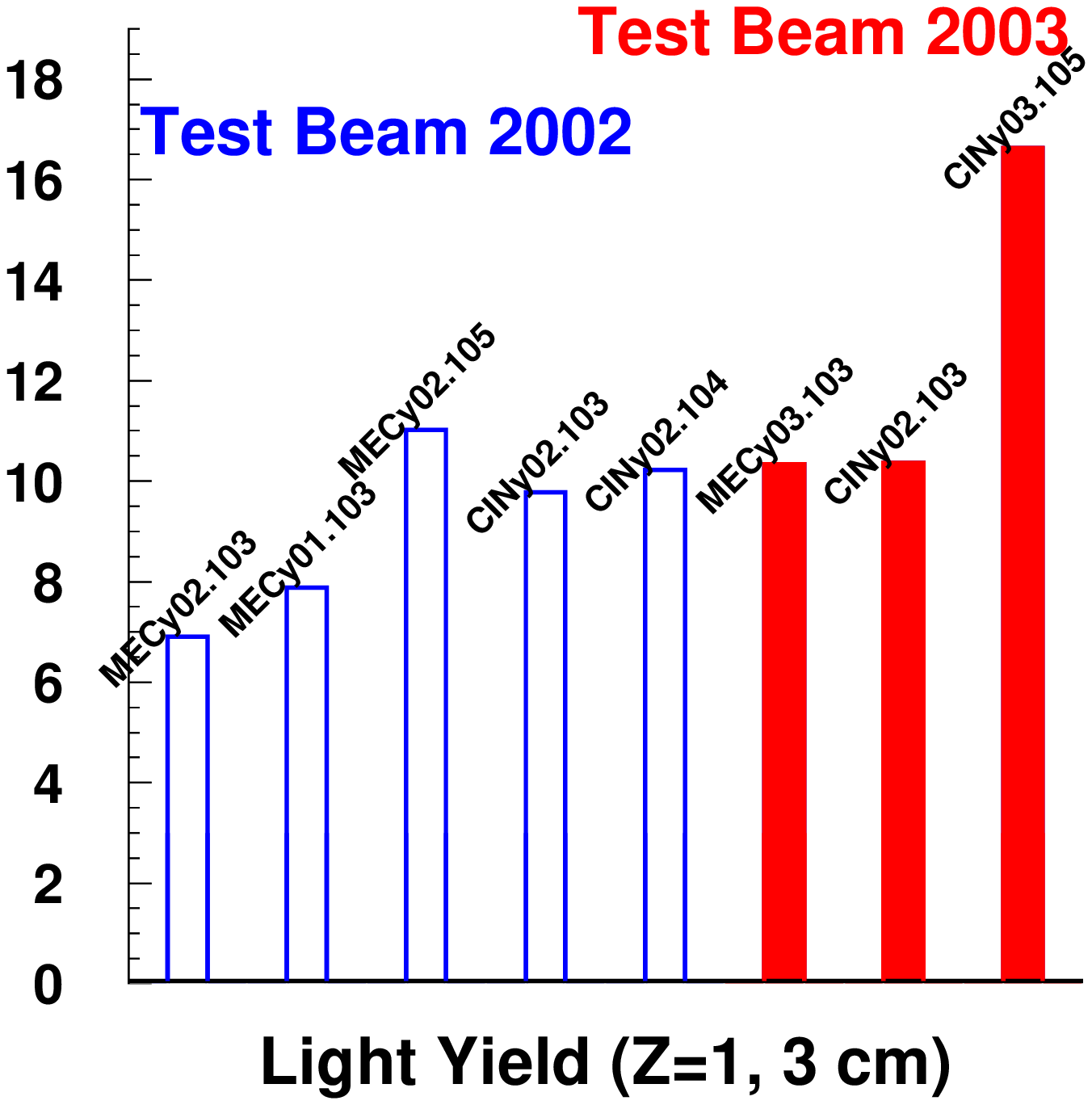} 
\vspace{-0.2cm}
   \caption{Comparison of the aerogel light yield for 3 cm thickness.   
     \label{fig:rich-yield}}
 \end{minipage} \hfill
 \begin{minipage}[b]{.32\linewidth}
   \psfrag{Zrec}{Z}
   \begin{overpic}[width=.95\textwidth]{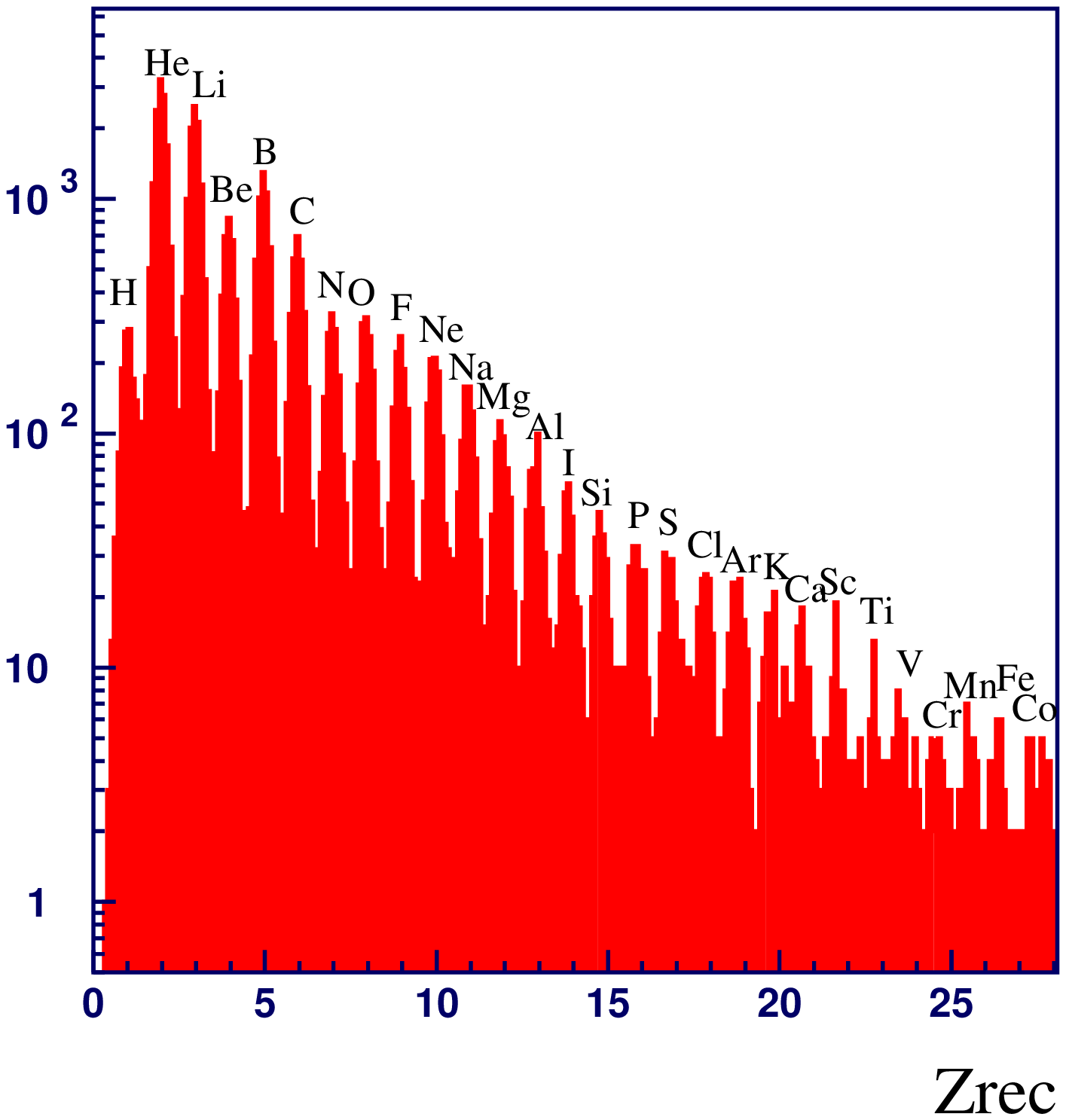}
   \put(50,80){CIN05}
   \end{overpic}
   \vspace{-0.2cm}
   \caption{Charge reconstructed ions from a 158 GeV/c primary In beam. 
     \label{fig:rich-chgrec}}
 \end{minipage} \hfill
 \begin{minipage}[b]{.32\linewidth}
   \centering\includegraphics*[width=0.9\textwidth,angle=0,clip]{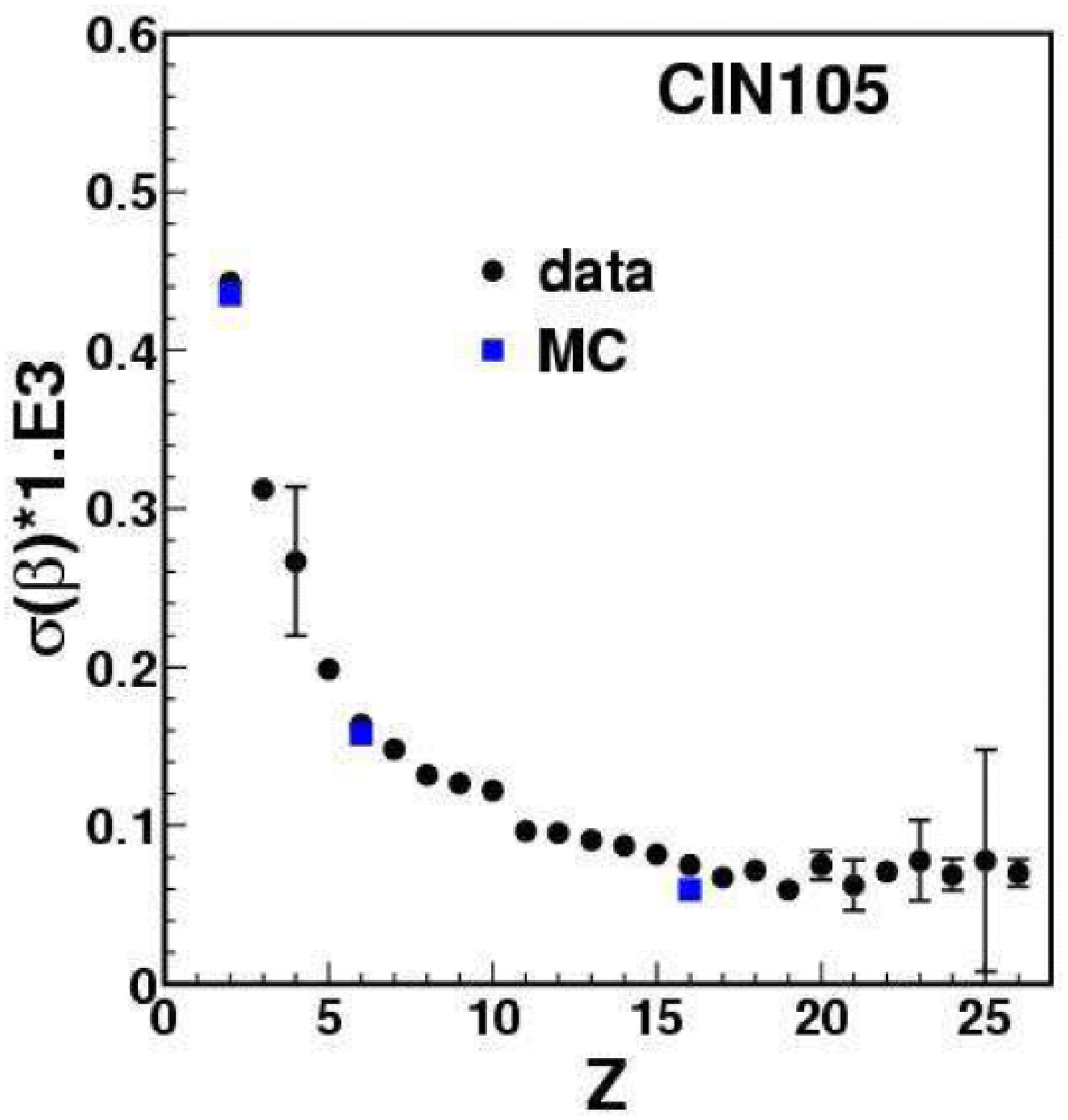} 
   \vspace{-0.2cm}
   \caption{Velocity accuracy dependence on the charge.  
     \label{fig:rich-betares}}
 \end{minipage}
\end{figure}

The evaluation of the aerogel samples in order to make a final radiator choice 
was one of the key issues of these tests.
Different production batches from two manufacturers, Matsushita Electric Co. (MEC) and   
Catalysis Institute of Novossibirsk (CIN) were analyzed.
The required criteria  for a good candidate were 
a high photon yield, in order to ensure a good ring reconstruction efficiency, and 
accurate $\beta$ and charge measurements.  
The aerogel light yield depends on the tile thickness and its optical properties, i.e refractive index and 
scattering effects (clarity).
Figure~\ref{fig:rich-yield} shows the normalized to 3 cm thickness light yield for the different 
aerogel samples tested in 2002 and 2003.
The highest signal comes from a CIN sample produced in 2003 with $1.05$ refractive index reflecting the very
good clarity ($\sim 0.0055~\mu$m$^4$/cm) of the aerogel batch.
The hydrophilic nature of this aerogel implies the sealing of the radiator container with a neutral 
gas like nitrogen in order to keep humidity below 50\%.
 
Reconstruction of velocity and charge were made with two independent methods~\cite{bib:rich-rec}. 
A charge resolution around $0.15$ is observed for low Z ions together with a systematic uncertainty, 
scaling with the charge, of 1.2\% due to non-uniformities.
Charge peaks up to iron were identified as shown in figure~\ref{fig:rich-chgrec}.
The velocity resolution scales with the detected signal ($\propto z^2$) as is shown on 
figure~\ref{fig:rich-betares}.
A relative accuracy $\Delta \beta/\beta \simeq 0.45~10^{-3}$ for heliums is obtained for the aerogel 
chosen (n=$1.05$). 
Similar clarity $1.03$ aerogels tested in 2003 had essentially the same resolution but lower photon yields.
Runs with a prototype mirror were also performed. 
The mirror reflectivity derived from data analysis is in good agreement with the design value.

\vspace{-0.5cm}
\section{Conclusions}
\vspace{-0.2cm}

A RICH detector is being constructed, and its assembling with the AMS spectrometer is scheduled for the beginning of 2006.
Cosmic muon and in-beam tests with fragmented ions validated the detector design and its goals; i.e, a singly charge 
resolution of 0.1\% and charge separation up to iron.
A refractive index $1.05$ aerogel was chosen for the radiator accommodating well both the demand for a large light yield and
good velocity resolution.   

\vspace{-0.5cm}
\section{Acknowledgments}
\vspace{-0.2cm}
We wish to thank the many organizations and individuals listed in  the
acknowledgments of reference~\cite{bib:ams}.

\end{document}